  \documentclass[prx,nofootinbib,superscriptaddress]{revtex4}
\usepackage{amssymb}  
\usepackage{xcolor}

\newcommand{\be}{\begin{equation}}
\newcommand{\ee}{\end{equation}}
\newcommand{\ba}{\begin{eqnarray}}
\newcommand{\ea}{\end{eqnarray}}

\def\a{\alpha}
\def\b{\beta}

\def\d{\delta}
\def\e{\epsilon}
\def\ve{\varepsilon}
\def\f{\phi}

\def\g{\gamma}
\def\h{\eta}
\def\j{\psi}
\def\k{\kappa}
\def\l{\lambda}
\def\m{\mu}
\def\n{\nu}

\def\r{\rho}
\def\s{\sigma}

\def\x{\xi}

\def\D{\Delta}
\def\F{\Phi}
\def\G{\Gamma}

\def\O{\Omega}

\def\S{\Sigma}

\def\cf{{\cal F}}

\def\co{{\cal O}}

\def\cs{{\cal S}}

\newcommand{\ov}{\overline}

\newcommand{\wh}{\widehat}

\newcommand{\aand}{\;\;\;\mbox{and}\;\;\;}

\newcommand{\pa}{\partial}

\newcommand{\pari}{\stackrel{{P}}\longrightarrow}
\def\sl#1{\rlap{\hbox{$\mskip 1 mu /$}}#1}
\def\Sl#1{\rlap{\hbox{$\mskip 3 mu /$}}#1}

\def\I{\leavevmode\hbox{\small1\kern-3.8pt\normalsize1}}

\begin{document}

\title{On the ultraviolet finiteness of parity-preserving $U(1) \times U(1)$ massive QED$_3$}

\author{W.B. De Lima}
\email{wellissonblima@cbpf.br}
\affiliation{Centro Brasileiro de Pesquisas F\'isicas (CBPF),\\
Rua Dr. Xavier Sigaud 150 - 22290-180 - Urca - RJ - Brazil.}

\author{O.M. Del Cima}
\email{oswaldo.delcima@ufv.br}
\affiliation{Universidade Federal de Vi\c cosa (UFV),\\
Departamento de F\'\i sica - Campus Universit\'ario,\\
Avenida Peter Henry Rolfs s/n - 36570-900 - Vi\c cosa - MG - Brazil.}
\affiliation{Ibitipoca Institute of Physics (IbitiPhys),\\
36140-000 - Concei\c c\~ao do Ibitipoca - MG - Brazil.}

\author{E.S. Miranda}
\email{emerson.s.miranda@ufv.br}
\affiliation{Universidade Federal de Vi\c cosa (UFV),\\
Departamento de F\'\i sica - Campus Universit\'ario,\\
Avenida Peter Henry Rolfs s/n - 36570-900 - Vi\c cosa - MG - Brazil.}

\begin{abstract}
The parity-preserving $U_A(1)\times U_a(1)$ massive QED$_3$ is ultraviolet finiteness -- exhibits vanishing $\beta$-functions, associated to the gauge coupling constants (electric and pseudochiral charges) and the Chern-Simons mass parameter, and all the anomalous dimensions of the fields -- as well as is parity and gauge anomaly free at all orders in perturbation theory. The proof is independent of any regularization scheme and it is based on the quantum action principle in combination with general theorems of perturbative quantum field theory by adopting the Becchi-Rouet-Stora (BRS) algebraic renormalization method in the framework of Bogoliubov-Parasiuk-Hepp-Zimmermann (BPHZ) subtraction scheme.
\end{abstract}
\maketitle

\section{Introduction}

The perturbative finiteness in quantum field theory, particularly in Chern-Simons models \cite{chern-simons} in three space-time dimensions, has drawn attention since the preliminary results at 1-loop order \cite{one-loop}, and afterwards at 2-loops \cite{two-loops}. At all orders in perturbation theory, pure non-Abelian Chern-Simons model in the Landau gauge exhibits ultraviolet finiteness \cite{chern-simons-landau-gauge}. However, even though coupled to bosonic and fermionic matter fields, non-Abelian Chern-Simons model in three-dimensional Riemannian manifolds still manifests at all radiative order vanishing $\b$-function associated to Chern-Simons coupling constant \cite{chern-simons-matter}. The massless $U(1)$ QED$_3$ exhibits ultraviolet and infrared perturbative finiteness, parity and infrared anomaly free at all orders \cite{masslessU1QED3}. Moreover, in opposition to some claims in the literature still now defending that parity could spontaneously be broken, even perturbatively, in massless $U(1)$ QED$_3$, known as parity anomaly, has already been discarded by the consistent and correct use of dimension regularization \cite{rao-delbourgo}, Pauli-Villars regularization \cite{pimentel}, algebraic renormalization in the framework of Bogoliubov-Parasiuk-Hepp-Zimmermann-Lowenstein (BPHZL) subtraction method \cite{masslessU1QED3}, and more recently through the Epstein-Glaser method \cite{epstein-glaser}. The exact quantum scale invariance in dimensional reduced to three dimensional space-time massless QED$_4$ models was investigated in \cite{dudal-mizher-pais}, and the gauge covariance of the massless fermion propagator was studied in quenched QED$_3$ \cite{quenchedQED3}. 
The massive $U(1)$ QED$_3$ can be odd (odd fermion families number) or even (even fermion families number) under parity symmetry. The parity-even massive $U(1)$ QED$_3$ is ultraviolet finite, the gauge coupling $\beta$-function and the anomalous dimensions of all the fields vanish, furthermore, is infrared and parity anomaly free at all orders \cite{massiveU1QED3}. Besides all the latter quantum field theory formal aspects, planar quantum electrodynamics (QED$_3$) has been demonstrated potential applications in condensed matter phenomena and low energy physics, on the other hand in early universe models and high energy physics as well. 

The main purpose in this work is to show the ultraviolet finiteness -- vanishing $\beta$-functions of both gauge couplings and all field anomalous dimensions -- at all orders in perturbation theory of the parity-even $U_A(1)\times U_a(1)$ massive QED$_3$ \cite{massiveU1U1QED3}, and the absence of any kind of anomaly, {\it e.g.} gauge and parity, as well. The proof is done by using the BRS (Becchi-Rouet-Stora) algebraic renormalization method in the framework of Bogoliubov-Parasiuk-Hepp-Zimmermann (BPHZ) subtraction scheme, which is based on general theorems of perturbative quantum field theory \cite{qap,brs,pigsor,zimm,low}, thus independent of any regularization scheme. Accordingly, the action of the model and its symmetries, the action for the gauge-fixing and the one which couples antifields to the nonlinear BRS transformations of the fields are established in Section \ref{II}. The extension of parity-even $U_{A}(1) \times U_{a}(1)$ massive QED$_3$ at the classical level to all orders in perturbation theory -- its perturbative quantization -- is arranged as follows. Prior the stability analysis of the classical action -- if the radiative corrections can be reabsorbed by a redefinition of the initial parameters of the model -- which is presented in Section \ref{IV}, in Section \ref{III} all potential anomalies are identified by means of the analysis of the Wess-Zumino consistency condition, in other words, solving the Slavnov-Taylor cohomology problem in the sector of ghost number one, in addition to, it is checked if the radiatively induced breakings might be fine-tuned by an appropriate choice of local non-invariant counterterms. Final comments and conclusions are left to Section \ref{V}.

\section{The model and its symmetries}
\label{II}

The action for the parity-even $U_{A}(1) \times U_{a}(1)$ massive QED$_3$ \cite{massiveU1U1QED3} is defined by:
\be
\S_{\rm inv}=\int{d^3 x} \biggl\{-{1\over4}F^{\m\n}F_{\m\n} - {1\over4}f^{\m\n}f_{\m\n} 
+ \m\e^{\m\r\n}A_\m\pa_\r a_\n + i {\ov\j_+} {\Sl D} \j_+ + i {\ov\j_-} {\Sl D} \j_- 
- m({\ov\j_+}\j_+ - {\ov\j_-}\j_-) \biggr\}~,\label{action_inv}
\ee
where ${\Sl D}\j_\pm \!\equiv\!(\sl\pa + ie\Sl{A} \pm ig\sl{a})\j_\pm$, $e$ (electric charge) and $g$ (pseudochiral charge) are the coupling constants with mass dimension $\frac{1}{2}$, and, $\m$ and $m$ $\m$ are mass parameters with mass dimension $1$. The field strengths, $F_{\m\n}=\pa_\mu A_\nu - \pa_\n A_\m$ and $f_{\m\n}=\pa_\mu a_\nu - \pa_\n a_\m$, correspond to the electromagnetic field ($A_\m$) and the pseudochiral gauge field ($a_\m$), respectively. $\j_+$ and $\j_-$ are two kinds of Dirac spinors where the $\pm$ subscripts are associated to their spin sign \cite{binegar}, and the gamma matrices are $\g^\m=(\s_z,-i\s_x,i\s_y)$. 

The action (\ref{action_inv}) was built up assuming invariance under parity, fixed posteriorly, and the gauge $U_A(1)\times U_a(1)$ transformations as follows:
\ba 
&&\delta\psi_+(x)=i[\theta_1(x) + \theta_2(x)]\psi_+(x)~,~~
\delta\psi_-(x)=i[\theta_1(x) - \theta_2(x)]\psi_-(x)~,\nonumber \\
&&\delta\ov{\psi}_+(x)=-i[\theta_1(x) + \theta_2(x)]\ov{\psi}_+(x)~,~~
\delta\ov{\psi}_-(x)=-i[\theta_1(x) - \theta_2(x)]\ov{\psi}_-(x)~,\nonumber \\
&&\delta A_{\mu}(x)=- \frac{1}{e}\partial_{\mu}\theta_1(x)~,~~
\delta a_{\mu}(x)=- \frac{1}{g}\partial_{\mu}\theta_2(x)~,\label{gaugetransf}
\ea
from which the BRS field transformations shall be defined. In view of a forthcoming quantization of the model (\ref{action_inv}), a parity-preserving gauge-fixing action is added, beyond that in order to follow the BRS procedure \cite{brs}, two sorts of Lautrup-Nakanishi fields ($b$ and $\pi$) \cite{lautrup-nakanishi}, ghosts ($c$ and $\x$) and  antighosts ($\ov{c}$ and $\ov{\x}$), the formers ($b$ and $\pi$) are indeed Lagrange multiplier fields fixing the gauge condition, have to be introduced. Therefore, so as to quantize the model, a gauge-fixing action, belonging to the class of covariant linear gauges \cite{thooft}, is assumed: 
\be
\S_{\rm gf} = \int {d^3x} \biggl\{ b \pa^\m A_\m + \frac{\a}{2}b^2 + \ov{c}\square c + 
\pi \pa^\m a_\m + \frac{\b}{2}\pi^2 + \ov{\x}\square \x \biggr\}~.\label{gauge_fixing}
\ee
Hereafter, the BRS transformations of the quantum fields are now defined by:
\ba
&s\j_+=i(c + \x)\j_+~,~~            s\ov{\j}_+=-i(c + \x)\ov{\j}_+~;& \nonumber \\
&s\j_-=i(c - \x)\j_-~,~~            s\ov{\j}_-=-i(c - \x)\ov{\j}_-~;& \nonumber \\
&\displaystyle sA_\m=-\frac{1}{e}\pa_\m c~,~~s c=0~;~~ \displaystyle sa_\m=-\frac{1}{g}\pa_\m \x~,~s\x=0~;& \nonumber \\
&\displaystyle s\ov{c}=\frac{b}{e}~,~~sb=0~;~~ \displaystyle s\ov{\x}=\frac{\pi}{g}~,~~s\pi=0~.&   \label{BRS}
\ea
Together with the parity-even action term, $\S_{\rm inv}+\S_{\rm gf}$, another parity-even action, $\S_{\rm ext}$, is introduced in order to control at the quantum level the renormalization of the nonlinear BRS transformations by coupling them to the antifields (BRS invariant external fields):
\be
\S_{\rm ext}\!\!=\!\!\!\int\!\!{d^3x}\biggl\{\! i\ov\O_+c_+\j_+ -i\ov{\O}_-c_-\j_- +ic_+\ov{\j}_+\O_+-ic_-\ov{\j}_-\O_- \! \biggr\}~, \label{action_ext}
\ee
where $c_+=c + \x$ and $c_-=c - \x$. It should be pointed out that in spite of the Faddeev-Popov ghosts be  massless, consequently serious infrared divergences could be stemmed from radiative corrections, nevertheless they decouple because of be free fields, there is no need to introduce Lowenstein-Zimmermann mass terms \cite{low} for them.
Now, the complete classical action to be perturbatively quantized reads
\be
\G^{(0)}=\S_{\rm inv}+\S_{\rm gf}+\S_{\rm ext}~.\label{action_total}
\ee

The propagators are the key ingredient on unitarity and spectral consistency analyses at tree-level of the model\footnote{The issues about unitarity, spectral consistency and two-particle scattering potentials have been discussed in \cite{moller}.}, even as in the calculation of ultraviolet and infrared dimensions of the fields. The tree-level propagators can be derived for all the quantum fields just by turning off the coupling constants ($e$ and $g$) and picking up the free part of the action $\S_{\rm inv}+\S_{\rm gf}$ ((\ref{action_inv}) and (\ref{gauge_fixing})). In that case all the propagators in momenta space are given by: 
\ba
 &&\D_{\ov{\j}\j}^{++}(k)= i\frac{{\sl k}-m}{k^2-m^2}~,~~ \D_{\ov{\j}\j}^{--}(k)\!= i\frac{{\sl k}+m}{k^2-m^2}~, \label{propk++--} \\
&&\D^{\m\n}_{AA}(k) =-i\bigg\{\frac{1}{k^2 - \mu^2}\left(\h^{\m\n} - \frac{k^\m k^\n}{k^2}\right) + \frac{\a}{k^2}\frac{k^\m k^\n}{k^2}\bigg\}~, \label{propkAA} \\     
&&\D^{\m\n}_{aa}(k) =-i\bigg\{\frac{1}{k^2 - \mu^2}\left(\h^{\m\n} - \frac{k^\m k^\n}{k^2}\right) + \frac{\b}{k^2}\frac{k^\m k^\n}{k^2}\bigg\}~, \label{propkaa} \\ 
&&\D^{\m\n}_{Aa}(k) =\frac{\mu}{k^2(k^2-\mu^2)}\e^{\mu\l\n}k_\l~, \label{propkAa} \\                            
&&\D^\m_{Ab}(k) =\D^\m_{a\pi}(k) =\frac{k^\m}{k^2}~, \label{propkAbapi} \\
&&\D_{\ov{c}c}(k) =\D_{\ov{\x} \x}(k) =-\frac{i}{k^2}~, \label{propkcc} \\ 
&&\D_{bb}(k) =\D_{\pi\pi}(k)=0~. \label{propkbbpipi}
\ea

It should be emphasized that the non decoupled propagators (\ref{propk++--})--(\ref{propkaa}) carrying physical degrees of freedom are all massive, so there were no infrared divergences that would arisen during the ultraviolet subtractions in BPHZ method, in other words from (\ref{propkAbapi})--(\ref{propkbbpipi}) it can be  concluded that all massless degrees of freedom -- so potential infrared divergences inducers -- do decouple. As a consequence, since no care about possible infrared divergences induced by ultraviolet subtractions is required, the ultraviolet divergences are the only that could spoiled the physical consistency of the model. Accordingly, the ultraviolet (UV) dimension of any fields, $X$ and $Y$, is established through the UV asymptotical behaviour ($d_{XY}$) of their propagator $\D_{XY}(k)$, namely $d_{XY}={\ov{\rm deg}}_{k}\D_{XY}(k)$, such that ${\ov{\rm deg}}_{k}$ provides the asymptotic power when $k\rightarrow \infty$. In addition to that, in the BPHZ renormalization method sometimes it is opportune to employ more UV subtractions than would indeed be necessary for convergence issue, establishing therefore bounds for UV subtractions of any Feynman graph or subgraph ($\g$) independent of either its detailed structure or its order. As a consequence, according to Zimmermann's convergence method \cite{zimm,low}, the UV dimension ($d$) of the fields, $X$ and $Y$, shall fulfill the inequality below:
\be
d_X + d_Y \geq d_{XY} + 3~, \label{uv-ir}
\ee
in such a manner that, together with the power-counting formula (\ref{powercounting}) presented farther, they set what kind of appropriate BPHZ ultraviolet subtractions have to be performed so as to get a finite integral associated to a particular graph or subgraph ($\g$) at some perturbative order.  

With the aim of determinate the UV dimensions of the vector fields, $A_\m$ and $a_\m$, and the spinor fields, $\j_+$ and $\j_-$, use has been made of the propagators (\ref{propk++--})--(\ref{propkAa}) together with the condition (\ref{uv-ir}), that results:
\ba
 &&d_{++}=-1~ \Rightarrow~ 2d_+\geq 2~ \Rightarrow~ d_+=1~;~~ 
   d_{--}=-1~ \Rightarrow~ 2d_-\geq 2~ \Rightarrow~ d_-=1~; \label{d+-}\\
 &&d_{AA}=-2~ \Rightarrow~ 2d_A\geq 1~ \Rightarrow~ d_A=\frac{1}{2}~;~~
   d_{aa}=-2~ \Rightarrow~ 2d_a\geq 1~ \Rightarrow~ d_a=\frac{1}{2}~; \label{dAAaa}\\
 &&d_{Aa}=-3~ \Rightarrow~ d_A + d_a\geq 0~, \label{dAa} 
\ea
where the latter constraint (\ref{dAa}){\color{blue}\footnote{The Zimmermann's inequality (\ref{uv-ir}) yields likewise the condition (\ref{dAa}) which is also satisfied by $d_A$ and $d_a$.}} -- stemming from the mixed propagator (\ref{propkAa}) and the UV bound (\ref{uv-ir}) -- is also fulfilled by UV dimensions $d_A$ and $d_a$ obtained in the former condition (\ref{dAAaa}). Additionally, it shall be called into question that, despite of the mixed propagator (\ref{propkAa}) carries no on-shell degrees of freedom \cite{massiveU1U1QED3,moller}, whenever it enters as an internal (off-shell) line of any 1-particle irreducible Feynman graph or subgraph ($\g$), thus as a virtual quantum, and the graph ($\g$) exhibits non-negative UV degree of divergence ($\d(\g)\geq 0$), BPHZ ultraviolet subtractions have to be performed. Notice however that, since $d_{Aa}<d_{AA}=d_{aa}$, internal lines containing the mixed propagator (\ref{propkAa}), instead of propagators (\ref{propkAA}) and (\ref{propkaa}), reduce the graph UV degree of divergence, as translated by the power-counting formula (\ref{powercounting}) presented in the following. Furthermore, concerning the fulfillment of the constraint (\ref{dAa}) by the UV dimensions $d_A$ and $d_a$, stemming from the mixed propagator (\ref{propkAa}), it might imply over-subtractions for some graph or subgraph ($\g$), either related to invariant counterterms or noninvariat counterterms, containing the propagator $\D^{\m\n}_{Aa}(k)$ as internal lines.

From the propagators (\ref{propkAbapi}) and the conditions (\ref{uv-ir}) and (\ref{dAAaa}), the UV dimensions of the Lautrup-Nakanishi fields, $b$ and $\pi$, can be fixed as:
\be
d_{Ab}=-1~ \Rightarrow~ d_A + d_b \geq 2~  \Rightarrow~ d_b=\frac{3}{2}~;~~
d_{a\pi}=-1~ \Rightarrow~ d_a + d_\pi \geq 2~ \Rightarrow~ d_\pi=\frac{3}{2}~. \label{dbpi} 
\ee
By considering the propagators (\ref{propkcc}), the UV dimensions of the Faddeev-Popov ghosts, $c$ and $\x$, and antighosts, $\ov{c}$ and $\ov{\x}$, are constrained by:
\be
d_{{\bar c}c}=-2~ \Rightarrow~ d_c + d_{{\bar c}} \geq 1~;~~ 
d_{{\bar \x}\x}=-2~ \Rightarrow~ d_\x + d_{{\bar \x}} \geq 1~. \label{constraint} 
\ee 
Furthermore, by fixing the BRS operator ($s$) as dimensionless and knowing that the coupling constants $e$ and $g$ have mass dimension $\frac{1}{2}$, from the conditions (\ref{constraint}), the UV dimensions of the Faddeev-Popov ghosts and antighosts result:
\be
d_c=0 \aand d_{\bar c}=1 ~;~~ d_\x=0 \aand d_{\bar \x}=1~.
\ee 
After all, from the antifields action ($\S_{\rm ext}$) together with the UV dimensions of all the quantum fields previously computed, it follows that:
\be
d_{\O_+}=2 \aand d_{\O_-}=2~.
\ee
Briefly, the UV dimension ($d$), the ghost number ($\F\Pi$) and the Grassmann parity ($GP$) of all fields are displayed in Table \ref{table1}. The statistics is defined in such a way that, the integer spin fields with odd ghost number and the half integer spin fields with even ghost number anticommute among themselves, in any other case the fields commute among themselves.

\begin{table}
\begin{center}
\begin{tabular}{|c|c|c|c|c|c|c|c|c|c|c|c|c|}
\hline
         &$A_\mu$ &$a_\mu$ & $\j_+$ & $\j_-$ &$c$  &${\ov c}$ &$b$          &  $\x$    & $\bar \x$ & $ \pi$         &$\O_+$   &$\O_-$  \\
\hline
$d$      &${1/2}$ &${1/2}$ &  1     &   1    & 0   &    1     &$\frac{3}{2}$&    0     &    1     &$\frac{3}{2}$   &    2    &     2   \\
\hline
$\F\Pi$  &   0    &   0    &  0     &   0    & 1   &  $-1$    &     0       &    1     &   $-1$   &      0         &   $-1$  &   $-1$   \\
\hline
$GP$     &   0    &   0    &  1     &   1    & 1   &    1     &     0       &    1     &   1      &      0         &     0   &    0     \\
\hline
\end{tabular}
\end{center}
\caption[]{The UV dimension ($d$), ghost number ($\F\Pi$) and Grassmann parity ($GP$).}\label{table1}
\end{table}

In a functional way, the Slavnov-Taylor identity expresses the BRS invariance of the action $\G^{(0)}$ (\ref{action_total}):
\be
\cs(\G^{(0)})=0~,\label{slavnovident}
\ee
where, acting on an arbitrary functional $\cf$, the Slavnov-Taylor operator $\cs$ read
\be
\cs(\cf) = \int{d^3 x} \biggl\{-{1\over e}{\pa}^\mu c {\d\cf\over\d A^\mu} + 
{b\over e} {\d\cf\over\d {\ov c}} - {1\over g}{\pa}^\mu \x {\d\!\cf\over\d a^\mu} + 
{\pi\over g} {\d\cf\over\d {\ov \x}} + 
{\d\cf\over\d \ov\O_+}{\d\cf\over\d \j_+} - {\d\cf\over\d \O_+}{\d\cf\over\d \ov\j_+} -  
{\d\cf\over\d \ov\O_-}{\d\cf\over\d \j_-} + {\d\cf\over\d \O_-}{\d\cf\over\d \ov\j_-} \biggl\}~.\label{slavnov}
\ee
and the linearized Slavnov-Taylor operator $\cs_\cf$ is given by
\ba
\cs_\cf &\!\!=\!\!&\int{d^3 x} \biggl\{-{1\over e}{\pa}^\mu c {\d\over\d
A^\mu} + {b\over e} {\d\over\d {\ov c}} +
-{1\over g}{\pa}^\mu \x {\d\!\over\d a^\mu} + {\pi\over g} {\d \over\d {\ov \x}} + \nonumber\\
&\!\!+\!\!&{\d\cf\over\d \ov\O_+}{\d\over\d \j_+} + {\d\cf\over\d \j_+}{\d\over\d \ov\O_+} - 
{\d\cf\over\d \O_+}{\d\over\d \ov\j_+} - {\d\cf\over\d\ov\j_+}{\d\over\d \O_+} - 
{\d\cf\over\d \ov\O_-}{\d\over\d \j_-} - {\d\cf\over\d \j_-}{\d\over\d \ov\O_-} + 
{\d\cf\over\d \O_-}{\d\over\d \ov\j_-} + {\d\cf\over\d\ov\j_-}{\d\over\d \O_-} \biggl\}~.\label{slavnovlin}
\ea
Thenceforward, it follows the nilpotency identities:
\ba
&&\cs_\cf\cs(\cf)=0~,~~\forall\cf~,\label{nilpot1} \\
&&\cs_\cf\cs_\cf=0~~{\mbox{if}}~~\cs(\cf)=0~.\label{nilpot3}
\ea
Particularly, the linearized Slavnov-Taylor operator $\cs_{\G^{(0)}}$ is nilpotent, namely $\cs_{\G^{(0)}}^2=0$, due to the fact that the action $\G^{(0)}$ (\ref{action_total}) fulfills the Slavnov-Taylor identity (\ref{slavnovident}). Moreover, the action of $\cs_{\G^{(0)}}$ upon the fields and the antifields (external sources) results
\ba
&&\cs_{\G^{(0)}}\f=s\f~,~~\f=\{\j_+,\ov\j_+,\j_-,\ov\j_-,A_\m,a_\m, b, c,{\ov c}, \pi, \x,{\ov \x}\}~,\nonumber\\
&&\cs_{\G^{(0)}}\O_+=-{\d\G^{(0)}\over\d\ov\j_+}~,~~\cs_{\G^{(0)}}\ov\O_+={\d\G^{(0)}\over\d\j_+}~,~~
\cs_{\G^{(0)}}\O_-={\d\G^{(0)}\over\d\ov\j_-}~,~~\cs_{\G^{(0)}}\ov\O_-=-{\d\G^{(0)}\over\d\j_-}~.\label{operation1}
\ea

Besides the Slavnov-Taylor identity (\ref{slavnovident}), the classical action $\G^{(0)}$ (\ref{action_total}) satisfies the gauge conditions, antighost equations and ghost equations as below:
\ba
&&\frac{\d \G^{(0)}}{\d b} = \pa^\m A_\m +\a b ~,~~
-i\frac{\d \G^{(0)}}{\d c} = i\Box\ov{c} + \ov{\O}_+\j_+ + \ov{\j}_+\O_+ - \ov{\O}_-\j_- - \ov{\j}_-\O_-~,~~
\frac{\d \G^{(0)}}{\d \ov{c}} = \Box c ~;\label{ghost_equation1} \\
&&\frac{\d \G^{(0)}}{\d \pi} = \pa^\m a_\m +\b \pi ~,~~
-i\frac{\d \G^{(0)}}{\d \x} = i\Box\ov{\x} + \ov{\O}_+\j_+ + \ov{\j}_+\O_+ - \ov{\O}_-\j_- - \ov{\j}_-\O_-~,~~ \frac{\d \G^{(0)}}{\d \ov{\x}} = \Box \x~.\label{ghost_equation2}  
\ea
Furthermore, the action $\G^{(0)}$ (\ref{action_total}) is invariant under the two rigid symmetries associated to $U_{A}(1) \times U_{a}(1)$:
\be
W_{\rm rig}^{e} \G^{(0)}=0 \aand W_{\rm rig}^{g} \G^{(0)}=0~,\label{rigidcond}
\ee
where the Ward operators, $W_{\rm rig}^{e}$ and $W_{\rm rig}^{g}$, read
\ba
W_{\rm rig}^{e} &\!\!=\!\!& 
\int{d^3 x}\biggl\{\j_+{\d\over\d \j_+} - \ov\j_+{\d\over\d \ov\j_+} + \O_+{\d\over\d \O_+} -
\ov\O_+{\d\over\d \ov\O_+} + \j_-{\d\over\d \j_-} - \ov\j_-{\d\over\d \ov\j_-} + \O_-{\d\over\d \O_-} - \ov\O_-{\d\over\d \ov\O_-}\biggr\}~,\label{wrigid_e}\\
W_{\rm rig}^{g}&\!\!=\!\!&
\int{d^3 x}\biggl\{\j_+{\d\over\d \j_+} - \ov\j_+{\d\over\d \ov\j_+} + \O_+{\d\over\d \O_+} -
\ov\O_+{\d\over\d \ov\O_+} - \j_-{\d\over\d \j_-} + \ov\j_-{\d\over\d \ov\j_-} - \O_-{\d\over\d \O_-} + \ov\O_-{\d\over\d \ov\O_-}\biggr\}~.\label{wrigid_g}
\ea

The $U_{A}(1) \times U_{a}(1)$ gauge invariant action $\G^{(0)}$ (\ref{action_total}) being even under the parity transformation ($P$) fixes its operation on the fields and antifields:
\ba
&& \j_+ \pari \j_+^P=-i\g^1\j_-~,~~ \j_- \pari \j_-^P=-i\g^1\j_+~,~~
\ov\j_+ \pari \ov\j_+^P=i\ov\j_-\g^1~,~~ \ov\j_- \pari \ov\j_-^P=i\ov\j_+\g^1~; \nonumber\\
&& \O_+ \pari \O_+^P=-i\g^1\j_-~,~~ \O_- \pari \O_-^P=-i\g^1\j_+~,~~
\ov\O_+ \pari \ov\O_+^P=i\ov\Omega_-\g^1~,~~ \ov\O_- \pari \ov\O_-^P=i\ov\Omega_+\g^1~; \nonumber\\
&& A_\mu \pari A_\mu^P=(A_0,-A_1,A_2)~;~~ \phi \pari \phi^P=\phi~,~~\phi=\{b, c, \ov{c}\}~; \nonumber\\
&& a_\mu \pari a_\mu^P=(-a_0,a_1,-a_2)~;~~ \chi \pari \chi^P=-\chi~,~~\chi=\{\pi, \x, \ov{\x}\}~.\label{parity_transformation}
\ea

Next, Section \ref{III} is devoted to seek for anomalies, {\it i.e.} classical symmetries broken at the quantum level. Although the gauge symmetry group $U_A(1)\times U_a(1)$ is Abelian, it is a non-semisimple Lie group, a priori the associated rigid invariance might be broken by radiative corrections, thus anomalous. Therefore, checking potential anomalies is a primary issue in comparison to the stability investigation presented in Section \ref{IV}

\section{The unitarity condition: in search for anomalies}
\label{III}

The multiplicative renormalizability, more precisely the stability condition, does not assure the extension of the classical model to quantum level, it still remains to guarantee the non existence of any gauge anomaly, {\it i.e.} electromagnetic and pseudochiral anomalies, and also the parity anomaly, once the latter is sometimes claimed in the literature as a typical anomaly of three dimensional space-times. 

The quantum vertex functional ($\G$) matches the tree-level action ($\G^{(0)}$) at zeroth-order in $\hbar$,
\be
\G = \G^{(0)} + {\co}(\hbar)~,\label{vertex}
\ee
shall fulfill the same conditions (\ref{ghost_equation1})--(\ref{rigidcond}) of the tree-level action.

Actually, as stated by the Quantum Action Principle \cite{qap}, the classical Slavnov-Taylor identity (\ref{slavnovident}) acquires a breaking at the quantum level:
\be
\cs(\G) =\D \cdot \G = \D + {\co}(\hbar \D)~,\label{slavnovbreak}
\ee
where the Slavnov-Taylor breaking $\D$ is an integrated local Lorentz invariant functional, with ghost number equal to $1$ and UV dimension bounded by $d\le\frac72$.

Taking into consideration the Slavnov-Taylor quantum identity ({\ref{slavnovbreak}), the nilpotency identity ({\ref{nilpot1}) applied to the quantum vertex functional, {\it i.e.}, $\cs_\G\cs(\G)=0$, and the equation  $\cs_{\G}=\cs_{\G^{(0)}} + {\co}(\hbar)$ obtained from (\ref{slavnovlin}) and (\ref{vertex}), this all together leads to the Wess-Zumino consistency condition for the quantum breaking $\D$:
\be
\cs_{\G^{(0)}}\D=0~,\label{breakcond1}
\ee
Moreover, in addition to (\ref{breakcond1}), calling into question the Slavnov-Taylor identity (\ref{slavnovident}), the gauge, antighost and ghost equations (\ref{ghost_equation1})--(\ref{ghost_equation2}), so as the rigid conditions (\ref{rigidcond}), it is verified that the Slavnov-Taylor quantum breaking ($\D$) also satisfies the constraints:
\be
{\d\D\over\d b}=\int d^3x \frac{\d\D}{\d c}={\d\D\over\d\ov c}=W_{\rm rig}^{e}\D=0 \aand 
{\d\D\over\d \pi}=\int d^3x \frac{\d\D}{\d \x}={\d\D\over\d\ov{\x}}=W_{\rm rig}^{g}\D=0~.\label{breakcond5}
\ee
At this point it shall be mentioned that, as far as rigid gauge invariance is concerned, since the symmetry group $U_A(1)\times U_a(1)$ is a non-semisimple Lie group, in principle rigid invariance could be broken at the quantum level, in other words, rigid symmetry might be anomalous. Nevertheless, none of both abelian factors are spontaneously broken as well as the conditions displayed in (\ref{rigidcond}), $W_{\rm rig}^{e} \G^{(0)}=0$ and $W_{\rm rig}^{g} \G^{(0)}=0$, express indeed the conservation of the electric ($e$) and the pseudochiral ($g$) charges, therefore the conditions exhibited in (\ref{breakcond5}), $W_{\rm rig}^{e}\D=0$ and $W_{\rm rig}^{g}\D=0$, are valid \cite{stora,kraus}.

Recalling again the Slavnov-Taylor condition (\ref{stabcond1}) satisfied by the counterterm (\ref{finalcount}), which is indeed a cohomology problem in the sector of ghost number zero, similarly in the sector of ghost number one, the cohomology problem is represented by Wess-Zumino consistency condition (\ref{breakcond1}). As a matter of fact, a general solution to the cohomology problem (\ref{breakcond1}) can ever be expressed as a sum of a trivial cocycle $\cs_{\G^{(0)}}{\wh\D}^{(0)}$, where ${\wh\D}^{(0)}$ has ghost number zero, and of nontrivial elements possessing ghost number one, ${\wh\D}^{(1)}$, lying in the cohomology of $\cs_{\G^{(0)}}$ (\ref{slavnovlin}):
\be
\D^{(1)} = {\wh\D}^{(1)} + \cs_{\G^{(0)}}{\wh\D}^{(0)}~,\label{breaksplit}
\ee
reminding that the Slavnov-Taylor quantum breaking $\D^{(1)}$ (\ref{breaksplit}) has to fulfill the constraints (\ref{breakcond1}) and (\ref{breakcond5}). It should be highlighted that the trivial cocycle $\cs_{\G^{(0)}}{\wh\D}^{(0)}$ can be incorporated order by order into the vertex functional $\G$, namely $\cs_{\G^{(0)}}(\G-{\wh\D}^{(0)}) = {\wh\D}^{(1)} + {\co}(\hbar \D^{(1)})$, as a non invariant integrated local counterterm, $-{\wh\D}^{(0)}$ -- a vanishing ghost number local field integrated polynomial. The linearized Slavnov-Taylor operator $\cs_{\G^{(0)}}$ (\ref{slavnovlin}) in combination with the Slavnov-Taylor quantum identity ({\ref{slavnovbreak}) results that the breaking $\D^{(1)}$ exhibits UV dimension bounded by $d\le{7\over2}$. It shall be stressed that since the insertion $\D^{(1)}$ stems from radiative corrections, it possesses a factor, $e^2$, $g^2$ or $eg$, at least, hence its effective UV dimension turns out to be bounded by $d\le{5\over2}$. 

Now, it has been verified that, as displayed in (\ref{breakcond5}), from the antighost equations (\ref{ghost_equation1})--(\ref{ghost_equation2}), the quantum breaking $\D^{(1)}$ (\ref{breaksplit}) fulfill the constraints: 
\be
\int d^3x \frac{\d\D^{(1)}}{\d c}=0 \aand \int d^3x \frac{\d\D^{(1)}}{\d \x}=0~,
\ee
then it follows that $\D^{(1)}$ reads
\be
\D^{(1)} = \int{d^3 x}\Bigl\{ {\cal K}^{(0)}_\mu\pa^\m c + {\cal X}^{(0)}_\m\pa^\m \x \Bigr\}~,\label{delta_1}
\ee
where ${\cal K}^{(0)}_\mu$ and ${\cal X}^{(0)}_\m$ are rank-$1$ tensors with zero ghost number and UV dimension bounded by $d\le{3\over2}$. Beyond that, the breaking $\D^{(1)}$ may be expressed by two linearly independent terms, where one is even under parity while the other is odd, thus ${\cal K}^{(0)}_\mu$ and ${\cal X}^{(0)}_\m$ can be written as 
\be
{\cal K}^{(0)}_\mu = \sum_{i=1} v_{k,i} {\cal V}^i_\mu + \sum_{i=1} p_{k,i} {\cal P}^i_\mu \aand 
{\cal X}^{(0)}_\mu = \sum_{i=1} v_{x,i} {\Upsilon}^i_\mu + \sum_{i=1} p_{x,i} {\Pi}^i_\mu~,
\ee 
with $v_{k,i}$, $p_{k,i}$, $v_{x,i}$ and $p_{x,i}$ being fixed coefficients to be further determined. 
Moreover, ${\cal V}^i_\mu$ and ${\Upsilon}^i_\mu$ are defined as vectors, while ${\cal P}^i_\mu$ and ${\Pi}^i_\mu$ as pseudo-vectors, in such a way that ${\cal V}^i_\mu\pa^\m c$ and ${\Pi}^i_\mu\pa^\m \x$ are parity-even, whereas ${\cal P}^i_\mu\pa^\m c$ and ${\Upsilon}^i_\mu\pa^\m \x$ are parity-odd, since $\pa^\m c$ is a vector and $\pa^\m \x$ a pseudo-vector. Taking into consideration that ${\cal K}^{(0)}_\mu$ and ${\cal X}^{(0)}_\mu$ have their UV dimensions given by $d\le{3\over2}$ and $\D^{(1)}$ must fulfill the conditions (\ref{breakcond1}) and (\ref{breakcond5}), it is verified that even though there are ${\cal P}^i_\mu$ and $\Upsilon^i_\mu$ surviving all of these constraints, namely, ${\cal P}^1_\mu=\e_{\m\r\n}\pa^\r A^\n$ and $\Upsilon^1_\mu=\e_{\m\r\n}\pa^\r a^\n$, their contributions in $\D^{(1)}$ (\ref{delta_1}) are all ruled out by partial integration, therefore effectively for the anomaly analysis purposes, $\{{\cal P}^i_\mu\}=\emptyset$ and $\{\Upsilon^i_\mu\}=\emptyset$, thereby leaving only the parity invariant part ($\D_{\rm even}^{(1)}$) of the Slavnov-Taylor quantum breaking $\D^{(1)}$ (\ref{delta_1}):
\be
\D_{\rm even}^{(1)} = \int{d^3 x}\Bigl\{ \sum_{i=1} v_{k,i}~{\cal V}^i_\mu\pa^\m c + 
\sum_{i=1} p_{x,i}~{\Pi}^i_\mu\pa^\m \x \Bigr\}~, \label{anomaly_even}
\ee
ending the proof that parity is not broken at the quantum level, there is no parity anomaly at all.
 
It still remains to verify if the parity-even breaking $\D_{\rm even}^{(1)}$ (\ref{anomaly_even}) is a genuine gauge anomaly or just a trivial cocycle that could be reabsorbed into the quantum action as noninvariant counterterms. However, it lacks to find the candidates for ${\cal V}_\m$ (vector) and ${\Pi}_\m$ (pseudo-vector) with UV dimensions given by $d\le{3\over2}$ provided that $\D_{\rm even}^{(1)}$ (\ref{anomaly_even}) satisfies the constraints (\ref{breakcond1}) and (\ref{breakcond5}). So, they read as follows:
\be
{\cal V}^{1}_\m = A_\m A^\n A_\n ~,~~ {\cal V}^{2}_\m = A_\m a^\n a_\n ~,~~ {\cal V}^{3}_\m = A_\n a^\n a_\m ~,~~
{\Pi}^{1}_\m = a_\m a^\n a_\n ~,~~ {\Pi}^{2}_\m = a_\m A^\n A_\n ~,~~ {\Pi}^{3}_\m = a_\n A^\n A_\m~, 
\ee
as a consequence the quantum breaking $\D_{\rm even}^{(1)}$ (\ref{anomaly_even}) becomes expressed by
\ba
\D_{\rm even}^{(1)} = \int{d^3 x}\Bigl\{\!\!\!\!\!\!\!
&&v_{k,1}A_\m A^\n A_\n \pa^\m c + v_{k,2}A_\m a^\n a_\n \pa^\m c + v_{k,3}A_\n a^\n a_\m \pa^\m c + \nonumber\\
&&+~p_{x,1}a_\m a^\n a_\n \pa^\m \x + p_{x,2}a_\m A^\n A_\n \pa^\m \x + p_{x,3}a_\n A^\n A_\m \pa^\m \x \Bigr\}~. 
\label{anomaly_even_1}
\ea  

At this moment, from the Wess-Zumino consistency condition (\ref{breakcond1}), whether any gauge anomaly thanks to radiative corrections shows up or not shall be demonstrated by analyzing the breaking $\D_{\rm even}^{(1)}$ (\ref{anomaly_even_1}), consequently if it could be written unambiguously as a trivial cocycle $\cs_{\G^{(0)}}{\wh\D}^{(0)}$, with ${\wh\D}^{(0)}$ being a zero ghost number integrated local monomials, the noninvariant counterterms, there is no gauge anomaly and the unitarity is guaranteed, otherwise, if there is at least one nontrivial element ${\wh\D}^{(1)}$ pertaining to the cohomology of $\cs_{\G^{(0)}}$ (\ref{slavnovlin}) in the sector of ghost number one, the gauge symmetry is anomalous and unitarity is definitely jeopardized. In order to give the sequel on the issue of gauge anomaly, it can be straightforwardly verified that:   
\ba
&&\cs_{\G^{(0)}}{\wh\D}^{(0)1} = \cs_{\G^{(0)}}\int{d^3 x}~A^\m A_\m A^\n A_\n = - \frac{4}{e}\int{d^3 x}~A_\m A^\n A_\n \pa^\m c ~, \label{cocycle_1}\\
&&\cs_{\G^{(0)}}{\wh\D}^{(0)2} = \cs_{\G^{(0)}}\int{d^3 x}~a^\m a_\m a^\n a_\n = - \frac{4}{g}\int{d^3 x}~a_\m a^\n a_\n \pa^\m \x ~, \\ 
&&\cs_{\G^{(0)}}{\wh\D}^{(0)3} = \cs_{\G^{(0)}}\int{d^3 x}~A^\m A_\m a^\n a_\n = - \frac{2}{e}\int{d^3 x}~A_\m a^\n a_\n \pa^\m c 
- \frac{2}{g}\int{d^3 x}~a_\m A^\n A_\n \pa^\m \x  ~, \\
&&\cs_{\G^{(0)}}{\wh\D}^{(0)4} = \cs_{\G^{(0)}}\int{d^3 x}~A^\m A_\n a^\n a_\m = - \frac{2}{e}\int{d^3 x}~A_\n a^\n a_\m \pa^\m c 
- \frac{2}{g}\int{d^3 x}~a_\n A^\n A_\m \pa^\m \x  ~.\label{cocycle_4}
\ea
Besides that, by taking into consideration the four trivial cocycles above (\ref{cocycle_1})--(\ref{cocycle_4}), it follows that the quantum breaking (\ref{anomaly_even}) might be rewritten as:
\ba
\D_{\rm even}^{(1)} &=& \cs_{\G^{(0)}} \left\{ \l_1 {\wh\D}^{(0)1} + \l_2 {\wh\D}^{(0)2} + \l_3 {\wh\D}^{(0)3} + \l_4 {\wh\D}^{(0)4} \right\} \nonumber\\
&=& \cs_{\G^{(0)}} \left\{ \l_1 \int{d^3 x}~A^\m A_\m A^\n A_\n + \l_2 \int{d^3 x}~a^\m a_\m a^\n a_\n + \l_3 \int{d^3 x}~A^\m A_\m a^\n a_\n + \l_4 \int{d^3 x}~A^\m A_\n a^\n a_\m \right\} \nonumber\\
&=& \int{d^3 x}\Bigl\{-\frac{4}{e}\l_1 A_\m A^\n A_\n \pa^\m c - \frac{2}{e}\l_3 A_\m a^\n a_\n \pa^\m c - \frac{2}{e}\l_4 A_\n a^\n a_\m \pa^\m c + \nonumber\\
&&\,~~~~~~~~~-\frac{4}{g}\l_2 a_\m a^\n a_\n \pa^\m \x - \frac{2}{g}\l_3 a_\m A^\n A_\n \pa^\m \x - 
\frac{2}{g}\l_4 a_\n A^\n A_\m \pa^\m \x \Bigr\}~,
\ea   
where it can be checked that
\be
v_{k,1}=-\frac{4}{e}\l_1 ~,~~ v_{k,2}=-\frac{2}{e}\l_3 ~,~~ v_{k,3}=-\frac{2}{e}\l_4 ~,~~ 
p_{x,1}=-\frac{4}{g}\l_2 ~,~~ p_{x,2}=-\frac{2}{g}\l_3 ~,~~ p_{x,3}=-\frac{2}{g}\l_4 ~,
\ee
thus finally demonstrating that the quantum breaking $\D_{\rm even}^{(1)}$ (\ref{anomaly_even}) is actually a trivial cocycle $\cs_{\G^{(0)}}{\wh\D}^{(0)}$:
\be
\D_{\rm even}^{(1)} = \cs_{\G^{(0)}}{\wh\D}^{(0)} = \cs_{\G^{(0)}} \left\{ \l_1 {\wh\D}^{(0)1} + \l_2 {\wh\D}^{(0)2} + \l_3 {\wh\D}^{(0)3} + \l_4 {\wh\D}^{(0)4} \right\}~.
\ee 
Therefore, as a final result, the integrated local monomials ${\wh\D}^{(0)1}$, ${\wh\D}^{(0)2}$, ${\wh\D}^{(0)3}$ and ${\wh\D}^{(0)4}$ can be absorbed as noninvariant counterterms order by order into the quantum action, {\it e.g.}, at $n$-order in $\hbar$:
\be 
\cs_{\G^{(0)}}(\G-\hbar^n{\wh\D}^{(0)}) \equiv \cs_{\G^{(0)}}\left(\G-\hbar^n\l_1 {\wh\D}^{(0)1}-\hbar^n\l_2 {\wh\D}^{(0)2}-\hbar^n\l_3 {\wh\D}^{(0)3}-\hbar^n\l_4 {\wh\D}^{(0)4}\right) = 0\hbar^n + {\co}(\hbar^{n+1})~, \label{a_final}
\ee
which concludes the proof on the absence of gauge anomaly, meaning that the $U_A(1)\times U_a(1)$ local symmetry is not anomalous at the quantum level.

In summary, it is finally proved that the parity-even $U_A(1)\times U_a(1)$ massive QED$_3$ \cite{massiveU1U1QED3} is free from any anomalies, stemming from either continuous or discrete symmetries, at all orders in perturbation theory. However, in the meantime in order to complete the full quantum level analysis, it still remains to demonstrate the multiplicative renormalizability of the model, which is presented in the following.

\section{The stability condition: in search for counterterms}
\label{IV}

The stability condition, {\it i.e.} the multiplicative renormalizability, is ensured if perturbative quantum corrections do not produce local counterterms corresponding to renormalization of parameters which are not already present in the classical theory, therefore thus those radiative corrections can be reabsorbed order by order through redefinitions of the initial physical quantities -- fields, coupling constants and masses -- of the theory. Consequently, so as to verify if the classical action $\G^{(0)}$ (\ref{action_total}) is stable under radiative corrections, it is perturbed by an arbitrary counterterm (integrated local functional) $\S^c$, namely $\G^\ve=\G^{(0)}+\ve \S^c$, such that $\ve$ is an infinitesimal parameter and the counterterm action $\S^c$ has the same quantum numbers as the tree-level action $\G^{(0)}$ (\ref{action_total}). Reminding the results of Section \ref{III} about the absence of any sort of anomaly, which means that all the classical symmetries are preserved at the quantum level, it can be straightforwardly concluded that the deformed action $\G^\ve$ satisfies all the conditions fulfilled by the classical action $\G^{(0)}$ (\ref{action_total}), this leads to the counterterm $\S^c$ be subjected to the following set of constraints:
\ba
&&\cs_{\G^{(0)}}\S^c=0~; \label{stabcond1}\\
&& W_{\rm rig}^{e}\S^c=0 ~,~~ W_{\rm rig}^{g}\S^c=0~; \label{stabcond2}\\
&&\frac{\d\S^c}{\d b}=\frac{\d\S^c}{\d c}=\frac{\d\S^c}{\d \ov{c}}=0 ~,~~ 
\frac{\d\S^c}{\d \pi}=\frac{\d\S^c}{\d \x}=\frac{\d\S^c}{\d \ov{\x}}=0~; \label{stabcond3}\\
&&\frac{\d \S^c}{\d\O_+}=\frac{\d \S^c}{\d\ov{\O}_+}=0 ~,~~ 
\frac{\d \S^c}{\d\O_-}=\frac{\d \S^c}{\d\ov{\O}_-}=0~. \label{stabcond4} 
\ea

The most general Lorentz invariant and vanishing ghost number field polynomial ($\S^c$) fulfilling the conditions (\ref{stabcond1})--(\ref{stabcond4}) with ultraviolet dimension bounded by $d\le3$, reads:
\be
\S^c = \int{d^3 x} \Bigl\{
\alpha_1 i{\ov\j}_+{\Sl D}\j_+ +  \alpha_2 i{\ov\j}_-{\Sl D}\j_- + 
\alpha_3 {\ov\j}_+\j_+ +  \alpha_4 {\ov\j}_-\j_- +\alpha_5 F^{\m\n}F_{\m\n} + 
\alpha_6 f^{\m\n}f_{\m\n}+ \alpha_7 \e^{\m\r\n}A_\m \pa_\r a_\n \Bigr\}~, \label{finalcount}
\ee
where $\alpha_i$ ($i=1,\ldots,7$) are parameters to be fixed later by normalization conditions. Beyond that, there are other constraints due to superrenormalizability and parity invariance. Since all quantum fields are massive no infrared divergences arise from the ultraviolet subtractions in the Bogoliubov-Parasiuk-Hepp-Zimmermann (BPHZ)  renormalization procedure, thereby there is no need to use the Lowenstein-Zimmermann subtraction scheme \cite{low} -- which explicitly breaks parity in three space-time dimensions \cite{masslessU1QED3,massiveU1QED3} while the infrared divergences are subtracted -- in order to subtract those infrared divergences. In this way, as a by-product the BPHZ method is an available parity-preserving subtraction procedure guaranteeing that at each loop order the counterterm ($\S^c$) shall be parity-even. The superrenormalizability of the model is responsible for the following coupling-constant-dependent power-counting formula:
\be
\d(\gamma)=3-\sum_{\F}d_\F N_\F-N_{Aa}-\frac{1}{2}N_e-\frac{1}{2}N_g~.  \label{powercounting}
\ee
is defined for the UV degree of divergence ($\d(\g)$) of a 1-particle irreducible Feynman diagram ($\g$),  whith $N_\F$ being the number of external lines of $\g$ associated to the field $\F$, $d_\F$ the UV dimension of $\F$ (see Table \ref{table1}), $N_{Aa}$ the number of internal lines of $\g$ associated to the mixed propagator $\D_{Aa}$(\ref{propkAa}), and $N_e$ and $N_g$ are the powers of the coupling constants $e$ and $g$, respectively, in the integral representing the Feynman graph $\g$. Thanks to fact that counterterms stem from loop corrections, thus they are at least of order two in the coupling constants, namely, $e^2$, $g^2$ or $eg$. Beyond that as previously mentioned, any topologically equivalent graphs with the same type of external legs, among those which have internal lines as the mixed propagator $\D_{Aa}$(\ref{propkAa}) instead of the pure ones $\D_{AA}$(\ref{propkAA}) or $\D_{aa}$(\ref{propkaa}), exhibit smaller UV degree of divergence due to the third term, $-N_{Aa}$, displayed in (\ref{powercounting}).

Accordingly, the effective UV dimension of the counterterm ($\S^c$) is bounded by $d\le2$, implying that, $\alpha_1=\alpha_2=\alpha_5=\alpha_6=0$. Furthermore, since a parity-even subtraction scheme is available thus the counterterm has to be parity invariant, resulting that $\alpha_3=-\alpha_4=\alpha$, and  the counterterm expressed by
\be
\S^c = \int{d^3 x}\Bigl\{\alpha({\ov\j}_+\j_+ - {\ov\j}_-\j_-)+ \alpha_7 \e^{\m\r\n}A_\m \pa_\r a_\n \Bigr\} = z_m m\frac{\pa}{\pa m}\G^{(0)} +z_\m \m\frac{\pa}{\pa \m}\G^{(0)}~,
\label{ct}
\ee
where $z_m=-\frac{\alpha}{m}$ and $z_\m=\frac{\alpha_7}{\m}$ are arbitrary parameters to be fixed order by order through the vacuum-polarization tensor and fermion self-energies normalization conditions:
\be
\frac{i}{6}\e_{\m\r\n} \frac{\pa}{\pa p_\r}\G^{\m\n}_{Aa}(p)\Big|_{p^2=\k^2} = \m ~,~~ \G_{{\ov\j}_+\j_+}(\sl{p})\Big|_{\sl{p}=+m} = 0 \aand \G_{{\ov\j}_-\j_-}(\sl{p})\Big|_{\sl{p}=-m} = 0~,
\ee 
with $\k$ being an energy scale. The counterterm (\ref{ct}) shows that, in principle, only the fermions mass $m$ and the Chern-Simons mass $\m$ shall acquire radiative corrections, implying that, at all orders in $\hbar$, the $\beta$-functions associated to the gauge coupling constants, $e$ and $g$, vanish, $\beta_e=0$ and $\beta_g=0$, respectively, the anomalous dimensions ($\g$) of any field as well. 

In time, a subtle property of the Chern-Simons piece of action in $\G^{(0)}$ (\ref{action_total}):
\be
\S_{\rm CS} = \m \int{d^3 x} ~\e^{\m\r\n}A_\m\pa_\r a_\n ~,\label{action_cs}
\ee
shall be put in evidence. The Chern-Simons action $\S_{\rm CS}$ (\ref{action_cs}) is not BRS local invariant, thus its invariance under BRS transformations is up to a surface term, {\it i.e.} a total derivative, 
\be
s\S_{\rm CS} = s \Bigl\{ \m \int{d^3 x} ~\e^{\m\r\n}A_\m\pa_\r a_\n \Bigr\} = 
-\frac{\m}{e} \int{d^3 x} ~\e^{\m\r\n}\pa_\m \left(c \pa_\r a_\n \right) ~,\label{s_action_cs}
\ee
suggesting a vanishing at the quantum level of the $\beta$-function \cite{CS-YMCS-BFYM,barnich} associated to the Chern-Simons mass parameter ($\m$), $\beta_\m=0$. In summary, the counterterm finally reads
\be
\S^c = \int{d^3 x}\Bigl\{\alpha({\ov\j}_+\j_+ - {\ov\j}_-\j_-)\Bigr\} = z_m m\frac{\pa}{\pa m}\G^{(0)} ~,
\label{ct_final}
\ee
yielding that all $\beta$-functions associated to the gauge coupling constants, electric charge ($e$) and pseudochiral charge ($g$), the Chern-Simons mass parameter ($\m$), and all anomalous dimensions ($\g$) of the fields, are vanishing, excepting the $\beta$-function corresponding to the fermions mass parameter ($m$). 

In summary, by taking the last result (\ref{ct_final}), the stability condition analysis, together with the former  one (\ref{a_final}), about the Wess-Zumino condition, ends the proof of vanishing $\beta$-functions associated to the gauge coupling constants ($e$ and $g$) and the Chern-Simons mass parameter ($\m$), and all anomalous dimensions ($\g$) of the fields, as well as the absence of parity and gauge anomaly at all orders in perturbation theory. Finally, as a by-product, it is demonstrated the all orders ultraviolet finiteness of the parity-even $U_A(1)\times U_a(1)$ massive QED$_3$ \cite{massiveU1U1QED3}. 


\section{Conclusion}
\label{V}

In conclusion, the parity-even $U_A(1)\times U_a(1)$ massive QED$_3$ \cite{massiveU1U1QED3} is free from any gauge anomaly and parity anomaly at all orders in perturbation theory. Beyond that, it exhibits vanishing $\beta$-functions associated to the gauge coupling constants ($e$ and $g$) and the Chern-Simons mass parameter ($\m$), and all the anomalous dimensions ($\g$) of the fields as well. The proof is independent of particular diagrammatic calculations or regularization schemes, since the BRS (Becchi-Rouet-Stora) algebraic renormalization method together with the BPHZ (Bogoliubov-Parasiuk-Hepp-Zimmermann) subtraction scheme \cite{qap,brs,pigsor,zimm,low} is grounded in the general theorems of perturbative quantum field theory. Furthermore, once the quantum perturbative physical consistency of the mass-gap graphene-like planar quantum electrodynamics has been proven from the results demonstrated here together with those presented in \cite{massiveU1U1QED3}, it should be newsworthy to deepen its analysis so as to apply in graphene-like electronic systems \cite{mudry}. As a final comment, the vanishing of all $\b$-functions associated to the electric charge ($e$), the pseudochiral charge ($g$) and the Chern-Simons mass parameter ($\m$) -- with the exception of that associated to the fermions mass parameter ($m$) -- foresees the independence of those system parameters with respect to the temperature, on the other hand for instance the mass-gap in graphene, which can be described by the fermions mass parameter, shall be temperature dependent.

\subsection*{Acknowledgements}

The authors thank the anonymous referee for helpful comments. O.M.D.C. dedicates this work to his father (Oswaldo Del Cima, {\it in memoriam}), mother (Victoria M. Del Cima, {\it in memoriam}), daughter (Vittoria), son (Enzo) and Glaura Bensabat. CAPES-Brazil and CNPq-Brazil are acknowledged for invaluable financial help.


\begin{references}

\bibitem{chern-simons} J.F. Schonfeld, Nucl. Phys. B185 (1981) 157;  R. Jackiw and S. Templeton, Phys. Rev. D23 (1981) 2291; S. Deser, R. Jackiw and S. Templeton, Ann. Phys. (NY) 140 (1982) 372 and Phys. Rev. Lett. 48 (1982) 975.

\bibitem{one-loop} R.D. Pisarski and S. Rao, Phys. Rev. D32 (1985) 2081; S. Deser, R. Jackiw and S. Templeton, Ann. Phys. (NY) 185 (1988) 406.

\bibitem{two-loops} G. Giavarini, C.P. Martin and F. Ruiz Ruiz, Nucl. Phys. B381 (1992) 222.

\bibitem{chern-simons-landau-gauge} A. Blasi and R. Collina, Nucl. Phys. B345 (1990) 472; F. Delduc, C. Lucchesi, O. Piguet and S.P. Sorella, Nucl.Phys. B346 (1990) 313; C. Lucchesi and O. Piguet, Nucl.Phys. B381 (1992) 281.

\bibitem{chern-simons-matter} O.M. Del Cima, D.H.T. Franco, J.A. Helay\"el-Neto and O. Piguet, J. High Energy Phys. 02 (1998) 002.

\bibitem{masslessU1QED3} O.M. Del Cima, D.H.T. Franco, O. Piguet and M. Schweda, Phys. Lett. B680 (2009) 108;  O.M. Del Cima, D.H.T. Franco and O. Piguet, Phys. Rev. D89 (2014) 065001.

\bibitem{rao-delbourgo} S. Rao and R. Yahalom, Phys. Lett. B172 (1986) 227; R. Delbourgo and A.B. Waites, Phys. Lett. B300 (1993) 241 and Austral. J. Phys. 47 (1994) 465. 

\bibitem{pimentel} B.M. Pimentel and J.L. Tomazelli, Prog. Theor. Phys. 95 (1996) 1217.

\bibitem{epstein-glaser} P.H. De Moura, O.M. Del Cima, D.H.T. Franco, L.S. Lima and E.S. Miranda, ``No parity anomaly in massless QED$_3$: an Epstein-Glaser approach'', in preparation.

\bibitem{dudal-mizher-pais} D. Dudal, A.J. Mizher and P. Pais, Phys. Rev. D99 (2019) 045017.

\bibitem{quenchedQED3} V.P. Gusynin, A.V. Kotikov and S. Teber, Phys. Rev. D102 (2020) 025013.

\bibitem{massiveU1QED3} O.M. Del Cima, Phys. Lett. B750 (2015) 1.

\bibitem{massiveU1U1QED3} O.M. Del Cima and E.S. Miranda, Eur. Phys. J. B91 (2018) 212.



\bibitem{qap} J.H. Lowenstein, Phys. Rev. D4 (1971) 2281 and Comm. Math. Phys. 24 (1971) 1; Y.M.P. Lam, Phys. Rev. D6 (1972) 2145 and Phys. Rev. D7 (1973) 2943; T.E. Clark and J.H. Lowenstein, Nucl. Phys. B113
(1976) 109.

\bibitem{brs} C. Becchi, A. Rouet and R. Stora, Comm. Math. Phys. 42 (1975) 127 and Ann. Phys. (N.Y.) 98 (1976) 287; O. Piguet and A. Rouet, Phys. Rep. 76 (1981) 1.

\bibitem{pigsor} O. Piguet and S.P. Sorella, {\em Algebraic Renormalization}, Lecture Notes in Physics, m28, Springer-Verlag (Berlin-Heidelberg), 1995.


\bibitem{zimm} W. Zimmermann, Comm. Math. Phys. 15 (1969) 208; P. Breitenlohner and D. Maison, Comm. Math. Phys. 52 (1977) 55.

\bibitem{low} J.H. Lowenstein and W. Zimmermann, Nucl. Phys. B86 (1975) 77; J.H. Lowenstein, Comm. Math. Phys. 47 (1976) 53; J.H. Lowenstein, {\em Renormalization Theory}, G. Velo, A.S. Wightman (Eds.), D. Reidel (Dordrecht-Holland), 1976.

\bibitem{binegar} B. Binegar, J. Math. Phys. 23 (1982) 1511.

\bibitem{lautrup-nakanishi} N. Nakanishi, Progr. Theor. Phys. 35 (1966) 1111; Progr. Theor. Phys. 37 (1967) 618; B. Lautrup, Mat. Fys. Medd. Dan. Vid. Selsk 35 (1967) No.11.

\bibitem{thooft} G. 't Hooft, Nucl. Phys. B33 (1971) 173; Nucl. Phys. B35 (1971) 167.

\bibitem{moller} O.M. Del Cima and E.S. Miranda, {\em On the emergence of attractive electron-electron interactions in $U(1) \times U(1)$ massive parity-even QED$_3$}, in progress.

\bibitem{stora} R. Stora, {\em Renormalization Theory}, G. Velo, A.S. Wightman (Eds.), D. Reidel (Dordrecht-Holland), 1976.

\bibitem{kraus} E. Kraus and K. Sibold, {\em Rigid Invariance in Gauge Theories}, proceedings of the Symposium at the {\em XXIst Conference International Colloquium on Group Theoretical Methods in Physics}, Group'21, ICGTMP'96 (Goslar-Germany), 1996, arXiv:hep-th/9703221v1. 

\bibitem{CS-YMCS-BFYM} O.M. Del Cima, D.H.T. Franco, J.A. Helay\"el-Neto and O. Piguet, J. High Energy Phys. 9802 (1998) 002; J. High Energy Phys. 9804 (1998) 010; Lett. Math. Phys. 47 (1999) 265.

\bibitem{barnich} G. Barnich, J. High Energy Phys. 9812 (1998) 003.

\bibitem{mudry} S. Ryu, C. Mudry, C.-Y. Hou and C. Chamon, Phys. Rev. B80 (2009) 205319.

\end{references}
\end{document}